\documentclass[referee]{raa}           
\usepackage{graphicx,times}
\usepackage{amssymb,amsmath}

\usepackage[a4paper=true,dvipdfm=true,pagebackref=true]{hyperref}
\hypersetup{pdftitle = The title of my PDF, pdfauthor = My name, pdfsubject= The subject, pdfkeywords = keyword1 keyword2 keyword3} 
\hypersetup{colorlinks = true, linkcolor = green, anchorcolor = red, citecolor = blue, filecolor = red, pagecolor = red, urlcolor = red}

\begin{document}

   \title{Measure Fiber Position Errors from Spectra Data
}

   \volnopage{Vol.0 (200x) No.0, 000--000}      
   \setcounter{page}{1}          

   \author{Jian-Jun Chen
      \inst{1,2,3}
   \and Zhong-Rui Bai
      \inst{1,2,3}
   \and A-Li Luo
      \inst{1,2}
   \and Yong-Heng Zhao
      \inst{1,2}
   }

   \institute{National Astronomical Observatories, Chinese Academy of Sciences,
             Beijing 100012, China; {\it jjchen@nao.cas.cn} \\
  \and
  Key Laboratory of Optical Astronomy, National Astronomical Observatories, Chinese Academy of Sciences,
	    Beijing 100012, China \\
  \and
  University of Chinese Academy of Sciences, Beijing 100049, China }


   \date{Received~~ month day; accepted~~~~month day}

\abstract{Precise fiber positioning is crucial to a wide field, multi-fiber spectroscopic survey like LAMOST. Nowadays, most position error measurements are based on CCD photographic and imaging processing techniques. Those methods only work for measuring errors orthogonal to the telescope optical axis, while there also lies errors parallel to the telescope optical axis, like defocusing, and error caused by the existing deviation angle between optical axes of a fiber and the telescope. Directly measuring two latter types of position errors is difficult for individual fiber, especially during observation. Possible sources of fiber position errors are discussed in brief for LAMOST. By constructing a model of magnitude loss due to the fiber position error for a point source, we propose an indirect method to calculate the total and systematic position errors for each individual fiber from spectra data. Restrictions and applications of this method are also discussed. 
\keywords{ techniques: multi-fiber --- fiber spectroscopy --- spectrophotometry  }
}

   \authorrunning{J.-J. Chen et al.  }            
   \titlerunning{Measure Fiber Position Errors from Spectra Data }  

   \maketitle
%
\section{Introduction}           
\label{sect:intro}

LAMOST is a special designed reflecting Schmidt telescope with 4000 fiber units mounted on the 5\dg field of view (FOV) focal plane, feeding targets light into 16 spectrographs  (Cui, X., et al.~\cite{cui12}). While 2dF and SDSS have achieved great successes in the last two decades, LAMOST is conducting the major multi-fiber spectroscopic survey (Zhao, G., et al.~\cite{zhao12}), and releasing spectra of more than one million targets each year. 

The large spectroscopic surveys primarily requires the signal-to-noise ratio (SNR) of observed spectra reaching certain criteria, while the SNR strongly depends on the proportion of light feeding into the effective aperture of each fiber from the target.  Considering the size of the LAMOST fiber is only 0.32 mm in diameter (corresponding to 3.3$''$), it is not easy to reach the position accuracy to about 1$''$ for 4000 fiber units in engineering, and even harder to maintain them in the entire cycle of years survey operation. Thus, routinely measuring the fiber position errors and keep the position accuracy is crucial to reach the SNR requirements of a survey like LAMOST. Moreover, besides the random position errors, the spatial distribution of fiber position errors on the focal plane may present a systematic pattern across the repeatedly observed fields over a period of time. This could bring up position depending selection effects and eventually jeopardize certain scientific goals of the survey. 

There are many sources contributing to fiber position errors. Newman (~\cite{newman02}) gave detailed discussion on the sources that impact the fiber position accuracy, especially, in the cases of SDSS and 2dF. There are three group of position errors according to Newman (~\cite{newman02}): the position errors orthogonal to the optical axis of telescope, the errors parallel to the optical axis, and the telecentric alignment errors, which represents an angle existing between the optical axes of the telescope and fibers. Although LAMOST does not adopt the magnetic puck-position system of 2dF, nor the drilled-plate system of SDSS, Newman's discussion is still suitable to LAMOST in general. Meanwhile, because LAMOST employs 4000 double revolving fiber position units on the focal plane to drive fibers parallel (Xing, X., et al.~\cite{xing98}, Cui, X., et al.~\cite{cui12}), there are some new sources that could affect the position accuracy, e.g. malfunction of stepping motors, machining and installation errors of motors and fiber units, etc.

Due to the importance of fiber position accuracy and the diversity of sources to position errors, many efforts have been carried out to measure and then to correct the position errors. So far most of such efforts on measurements are based on CCD photographic and image processing techniques. LAMOST routinely takes photographic measurement to calibrate the focal surface coordinates and check the working condition of fiber units. In practice, the measurement for LAMOST fiber positioning is conducted about every 3 months in order to test the precision of the fiber position system. A series of CCD photos are taken and processed for the focal plane, while the fiber-heads have been illuminated from spectrograph-ends and positioned to multiple testing positions to calibrate the focal plane coordinates with the help of a standard spots array (Cui, X., et al.~\cite{cui12}). The malfunctioning fiber units could also be found out and replaced after this process. 

Recently, modified CCD photographic methods were continually proposed, either to achieve higher precision (Gu, Y., et al.~\cite{gu12}), or to cut down the spending time of measurement in order to have near real-time measuring and feedback during the observation (Wang,M., et al.~\cite{wang12}). But all the photography based methods are only able to measure the position errors orthogonal to the optical axis, with little help to deal with the errors parallel to the optical axis and the error of fiber telecentric alignment.

In this paper, a new approach is proposed to measure the total position errors, including orthogonal, parallel and telecentric alignment errors. This approach is based on a model describing the magnitude loss of a point source due to position errors in various seeing conditions. While the light of a targeted point source and the light of sky background identically falls into a fiber's aperture, in principle the sky brightness magnitude can be calculated by providing the magnitude of point source from the input catalog, and the target and sky flux from the observed spectrum. Actually the sky brightness from this solution is varied along with the value of position errors, i.e. the bigger position errors, the solution gives the larger sky brightness. Of course that is not true, and it is because the flux corresponding to the magnitude from the input catalog does not fully fall into the fiber aperture due to the position errors. Given a true sky brightness magnitude, the difference between the true and calculated sky brightness represents the magnitude loss due to fiber position errors, and the quantity of position errors then can be solved from the mentioned model. 

In the rest of this paper, Section~\ref{sect:position-errors} briefly explores the sources of fiber position errors, particularly, in the case of LAMOST fiber position system. Section~\ref{sect:model} introduces a concept of equivalent position error. It only has nonzero orthogonal component, and has the identical magnitude loss to that caused by all three groups of position errors. Then a model is presented to quantitatively describe the correlation between the magnitude loss of a point source and the equivalent position error. Section~\ref{sect:skybright} describes the procedure that deduces the quantity of equivalent position error by comparing the true sky brightness and the sky brightness calculated from input catalog and spectra flux. In section~\ref{sect:discussion}, several aspects of this measurement are discussed, including the influence of atmospheric transmittance, the possible errors in this method, and the comparison to SNR and photographic measurements. 


\section{Sources of Position Errors of LAMOST}
\label{sect:position-errors}

Newman (~\cite{newman02} ) gave a comprehensive description to the sources of fiber-to-image position mismatching in the multi-fiber feeding spectroscopic telescope, particularly, in the cases of SDSS and 2dF. LAMOST fiber position system takes advantage in the efficiency of fiber positioning. It is able to accomplish a procedure of fiber reconfiguration in minutes with the ability to position 4000 fibers simultaneously. While the spherical focal surface actually is composed by the  head-ends of 4000 individual fiber units, besides the sources of position errors listed in Newman(~\cite{newman02}), the fiber position precision of LAMOST strongly depends on the fabrication and installation accuracy of these 4000 fiber units. Maintaining the accurate position system depends on the working conditions of these fiber units.

Table~\ref{tab1} gives a summary of the sources of the position errors and the corresponding measurement/correction dealing with position errors applied in the operation of LAMOST.

\begin{table}
\begin{center}
 \caption[]{Position Error Sources and Measurement/Correction Applied to LAMOST\label{tab1}}

 \begin{tabular}{p{8cm}p{8cm}}
  \hline\noalign{\smallskip}
   \hspace{0.5cm}{\bf{Position Error Sources}} & {\bf{Measurement/Correction Applied to LAMOST}} \\
  \hline\noalign{\smallskip}
   {\bf{ Errors orthogonal to optical axis}} &        \\ 
   \hspace{0.5cm}Astrometry & Input catalog and guiding system            \\ 
   \hspace{0.5cm}Aberration, parallax and proper motion & Input catalog and guiding system\\
   \hspace{0.5cm}Conversion to focal surface coordinates & Calibration CCD and "fiber scan" (Cui, X., et al. ~\cite{cui12} ) \\
   \hspace{0.5cm}Fiber and fiber unit mounting & Calibration CCD images \\
   \hspace{0.5cm}Temporal variation in image scale & Guiding CCD images and adaptive optical \\
   \hspace{0.5cm}Collimation and field rotation & Guiding CCD images and adaptive optical \\
   \hspace{0.5cm}Atmospheric distortion and guiding & Guiding CCD images/guiding system \\
   \hspace{0.5cm}Atmospheric differential refraction & Restricted observation area (Donnelly, R., et al. ~\cite{Donnelly89} )\\
  \hspace{0.5cm}Telescope Pointing & Guiding system \\
  \hspace{0.5cm}Stepper motor malfunction & Motor controller feedback and software block \\
   {\bf{ Errors parallel to optical axis}}    &                         \\
    \hspace{0.5cm}Shape of focal surface & manufacturing/installation accuracy of fiber units        \\
   \hspace{0.5cm}Focus errors during observation & Guiding CCD images/adaptive optical \\
   {\bf{Telecentric alignment error }}    &      \\
    \hspace{0.5cm}Angle between fiber axis and telescope optical axis & Installation accuracy of fiber units \\

  \noalign{\smallskip}\hline
\end{tabular}
\end{center}
 \end{table}

\section{Model of Magnitude Loss due to Position errors}
\label{sect:model}
For a point source, the fiber-to-image position mismatch certainly causes flux loss of this target, thus, the spectrophotometric magnitude loss. Conversely, the magnitude loss also leads to measuring the total position errors of the fiber targeting a point source. While the total position errors include orthogonal, parallel, and telecentric alignment errors, a conception of equivalent position error is introduced in this paper for constructing the magnitude loss model. The equivalent position error is defined as setting parallel and telecentric alignment components to zero, and only having orthogonal component left nonzero, meanwhile it causes the effects of magnitude loss equivalent to that caused by total position errors. 

The image profile of a point source on the focal surface is quite complex, while it has been convolved with the system response function of the telescope, the turbulence of atmosphere, random motion of guiding adjustment, and then integrated over exposure time. In this paper, the image profile of a point source is adequately modelled by a normalized two dimensional Gaussian (Brodie, J., et al.~\cite{Brodie88}) with known  $\sigma=W/(2\sqrt{ln4})$: 

\begin{equation}
\label{eq1}
 f(x,y)=\frac{1}{2\pi\sigma^2}e^{-\frac{(x-\Delta x)^2 + y^2 }{2\sigma^2}}
\end{equation}
Where $\Delta x$ is the equivalent position error.

The quantity of the width of the point spread function, $W$, is given by measuring FWHM of the star spots  on the guiding CCD images. It is affected by both dome and atmospheric seeing, and any systematic mis-focus across the focal plane. In this paper, a constant seeing disk across the focal plane is assumed. While four guiding CCD cameras are mounted in a square on the LAMOST focal plane (Cui, X., et al.~\cite{cui12}), it is easy to find out if this assumption is satisfied by checking the variation of $W$ quantities among four guiding CCDs images. 

Considering the diameter of fiber is 3.3$''$, given $\Delta x$ and $W$ , the flux $F(\Delta x)$ falling into the fiber aperture is the integration:

\begin{equation}
\label{eq2}
 F(\Delta x)=\int_{-1.65}^{1.65}\int_{-\sqrt{1.65^2-x^2}}^{\sqrt{1.65^2-x^2}} f(x,y) dxdy
\end{equation}

Given $\Delta x_{0}=0$ and a set of $\Delta x_{i} > 0$, the corresponding magnitude loss $m_{a}$ is:

\begin{equation}
\label{eq3}
m_{a}(i)=mag(\Delta x_{i})-mag(\Delta x_{0})=-2.5log\frac{F(\Delta x_{i})}{F(\Delta x_{0})}
\end{equation}
Where $F(\Delta x_{0})=F(\Delta x=0)$ is the flux into the fiber aperture when the equivalent position error $\Delta x =0$.

\begin{figure}
   \centering
   \includegraphics[width=10.0cm, angle=0]{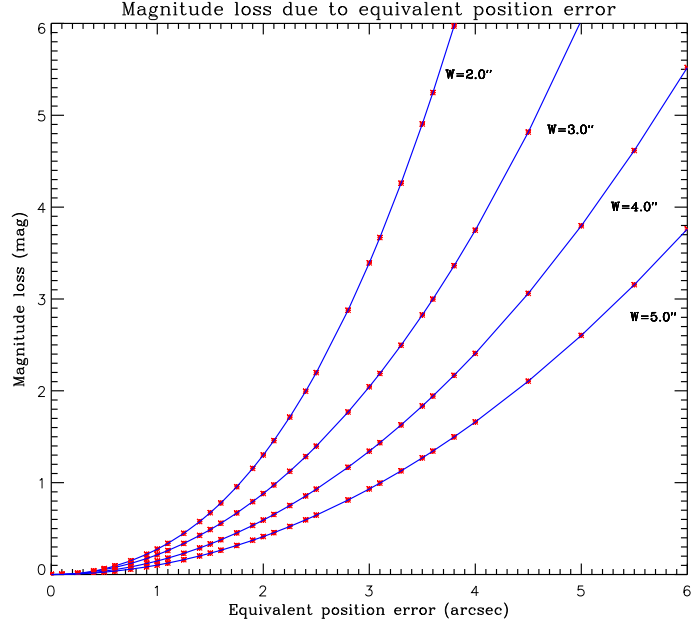}
   \caption{The model of magnitude loss ($m_{a}$) due to the equivalent position error ($\Delta x$) for a point source. The various seeing condition ($W$) gives different polynomial fitted curves. } 
   \label{Fig1}
   \end{figure}

Fig.\ref{Fig1} gives the data points of a point source's magnitude loss $m_{a}(i)$ corresponding to the equivalent position error $\Delta x_i$, and the polynomial fitted curves at various seeing ($W$) conditions.

\section{Sky Brightness and Equivalent Position Error}
\label{sect:skybright}
The flux feeding into the aperture of the $i$th fiber is actually composed of two components: the flux from the targeted object and the flux from the sky background. These two components fall into exactly the same aperture and identically convolve with the total system response of the telescope and instruments. The sky light is the mixture of airglow, background light from faint celestial objects, zodiac light, and ground pollution light, etc. (Roach, F.,~\cite{Roach64}, and Gustafson, B., et al.~\cite{Gustafson07}) Some of them are from the outside of atmosphere, similar to the light from the targets, and some are not. Therefore, the magnitude of sky brightness $m_{sky}$ can be calculated from the equation:

\begin{equation}
\label{eq4}
m_{obs}(i)-m_{sky}(i)=-2.5log\frac{flux_{obs}(i)}{flux_{sky}(i)}+\Delta m_{atm}(i)
\end{equation}
Where $flux_{obs}(i)$ is the sky-subtracted flux of a point source targeted by the $i$th fiber; $flux_{sky}(i)$ is the flux of sky background light feeding into the $i$th fiber, which in practice is composed of the spectra of nearby sky sampling fibers; $m_{obs}(i)$ and $m_{sky}(i)$ are their corresponding magnitudes; and $\Delta m_{atm}(i)=-2.5log(\Delta flux_{obs,atm}(i)/\Delta flux_{sky,atm}(i))$ represents the magnitude difference caused by the various atmosphere thickness passed by target and sky light respectively. 

For each target, LAMOST input catalog provides the photometric magnitude $m_{obj}$ retrieved from high precise multi-band photometric catalogs, like SDSS, PanStarrs, and Xuyi Antigalactic Center photometric survey, etc. (Zhao, G., et al. ~\cite{zhao12}) Considering the size of seeing disks, only part of target's light falls into the fiber aperture of 3.3$''$ in diameter, the corresponding magnitude $m_{obj}'$ is not equal to $m_{obj}$:

\begin{equation}
\label{eq5}
 m_{obj}=m_{obj}'-2.5log\frac{\iint_{-\infty}^{\infty}f(x,y)dxdy}{\iint_{\sqrt{x^2+y^2}<1.65}f(x,y)dxdy}=m_{obj}'-2.5log(1/F(\Delta x_{0}))
\end{equation}

The night sky, especially the dark night sky, has been taken as the uniform source for the flat field exposures in observational practice for a long time. Uniformity of dark sky at the zenith is near perfect in a clear, dark night. The relative gradient slowly degrades to one percent per degree to zenith angle about 50\dg, and degrades further to about 2\% per degree until zenith angle close to 70\dg ~(Chromey. F., et al, \cite{Chromey96}). 

Therefore the sky magnitude differential among the fibers in one exposure would be very small if this exposure is taken in the clear and dark night, and the zenith angle is limited within 50\dg. In this condition, the magnitude gradient of sky brightness among the 4000 fibers across 5\dg ~FOV is less than 0.05 mag. This number is negligible, comparing to the value of most concerned magnitude loss in the Fig.\ref{Fig1}, where equivalent position error is larger than 1.0$''$. 

The value of $m_{atm}$ is depending on airmass (Donnelly, R., et al. ~\cite{Donnelly89}). Near the zenith angle of 30\dg, the difference of air mass across 5\dg ~FOV is about 5\%, corresponding to the maximal difference of 0.05 mag among 4000 fibers. When zenith angle increases to 50\dg, the air mass differential changes to about 10\% across 5\dg, corresponding the maximum of about 0.1 mag among 4000 fibers. So there is a need to consider $m_{atm}$ differential when zenith angle is larger than 30\dg. If we select an exposure with the pointing of zenith angle less than 30\dg, ignoring $m_{sky}$ and $m_{atm}$ differential among fibers in 5\dg ~FOV, we have $m_{sky}(i)=m_{sky}$ and $\Delta m_{atm}(i)=m_{obs,atm}(i)-m_{sky,atm}(i)=\Delta m_{atm}$. Note that here $m_{obs}=m_{obj}'+m_{a}$ by the definition and let the pseudo sky brightness $m_{sky}^{'}=m_{sky}-\Delta m_{atm}$, the Eq. \ref{eq4} can be rewritten as :

\begin{equation}
\label{eq6}
I(i)=m_{sky}^{'} - m_{a}(i)=m_{obj}(i)+2.5log\{\frac{flux_{obs}(i)}{flux_{sky}(i)}\cdot\frac{1}{F(\Delta x_{0})}\}
\end{equation}

Where $I(i)=m_{sky}^{'} - m_{a}(i)$ is defined as the implied sky brightness.

\begin{figure}[h]
  \begin{minipage}[t]{0.495\linewidth}
  \centering
   \includegraphics[width=67mm,height=56mm]{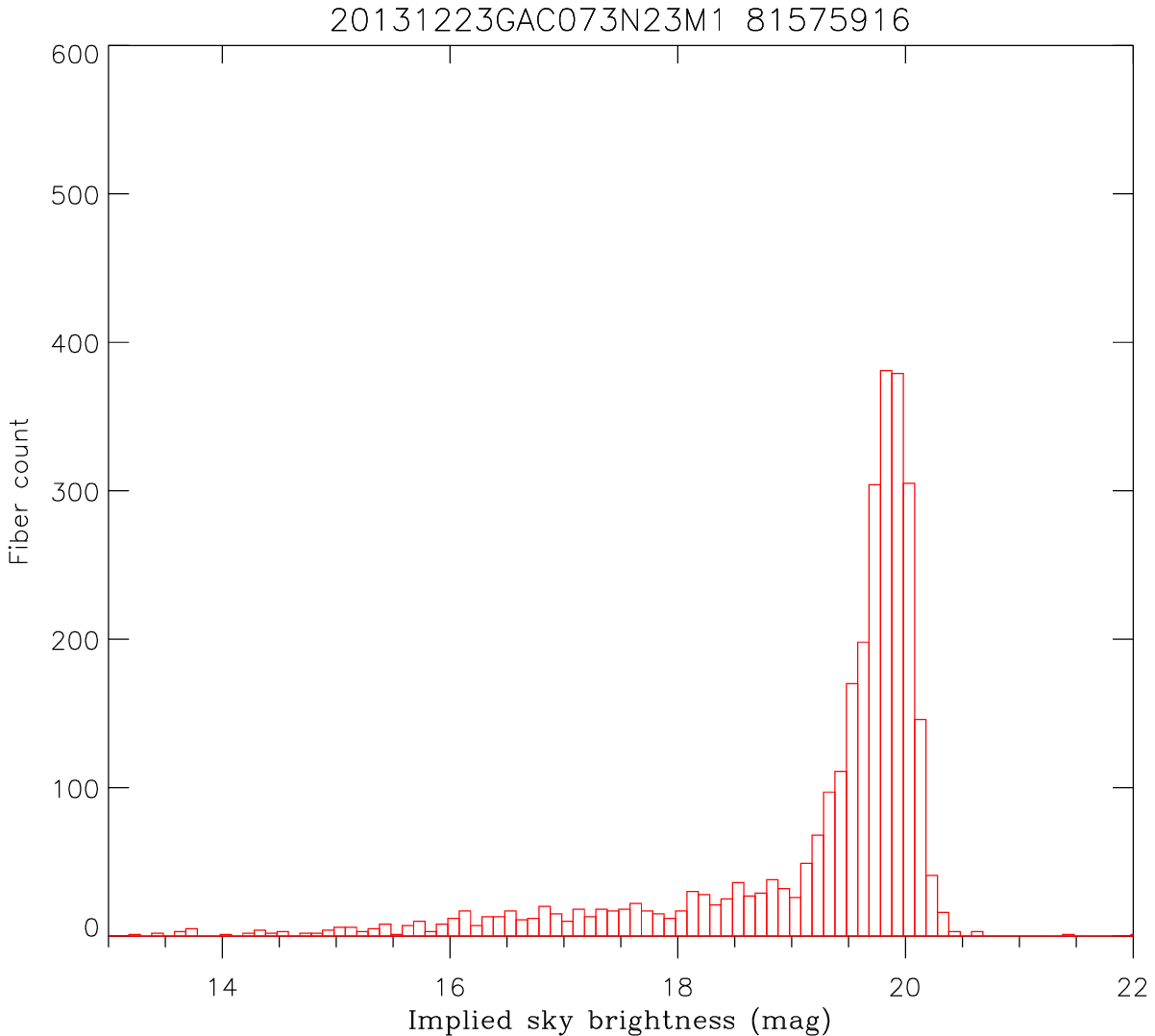}
   \caption{{\small Histogram for implied sky brightness $I(i)=m_{sky}^{'} - m_{a}(i)$ of one exposure with a seeing disk $W$ of 2.6$''$. The peak of the distribution is 19.85 mag,  The uniform sky brightness is estimated to be 20.5 mag (by maximum) or 20.38 mag (by $3\sigma$ cut) on this plot, where $\sigma$ is the variation of the right side half Gaussian from the peak.} }
  \label{fig2}
  \end{minipage}%
  \begin{minipage}[t]{0.495\textwidth}
  \centering
   \includegraphics[width=72mm,height=60mm]{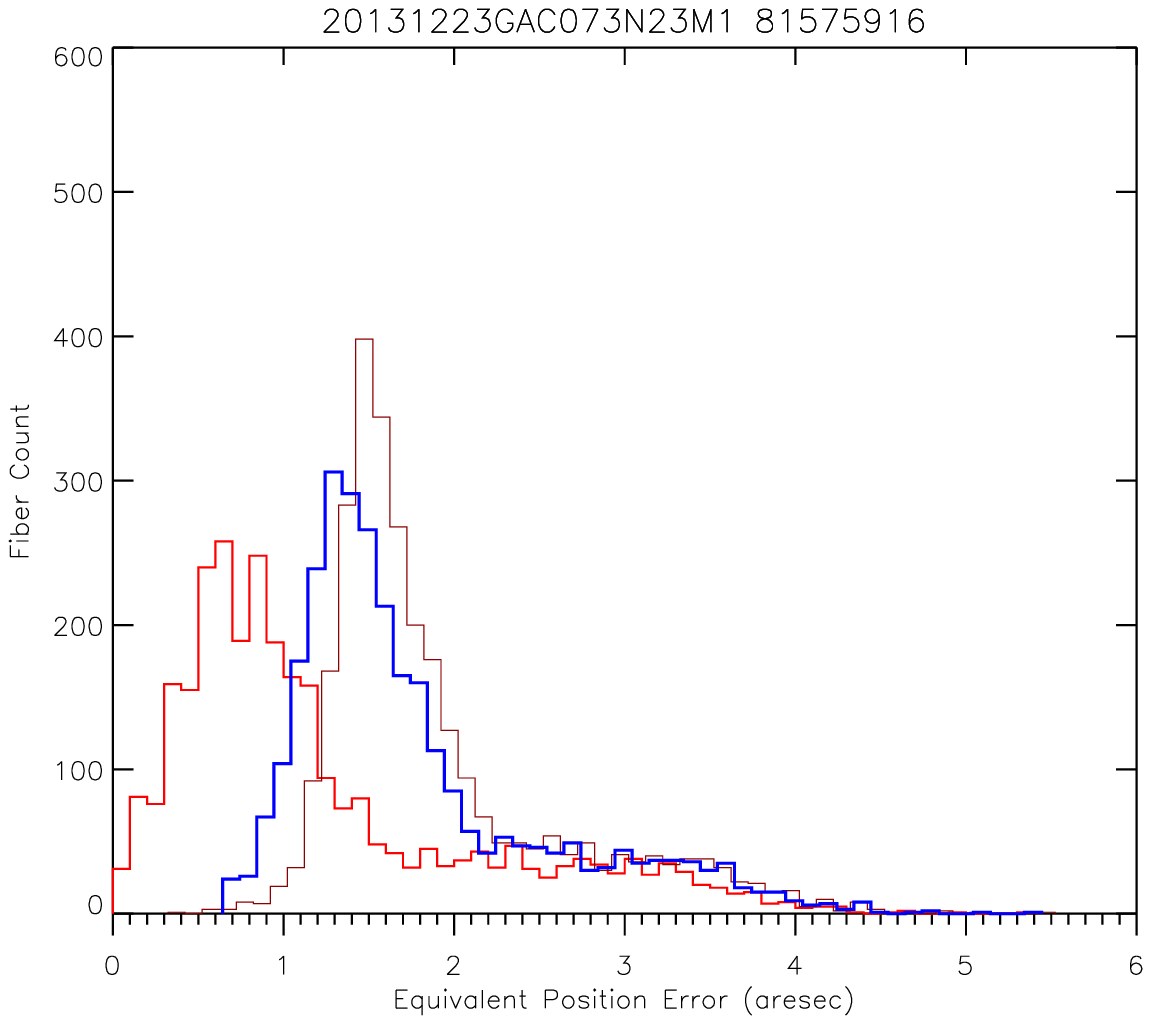}
  \caption{{\small The value of equivalent position error distributions is solved by the error model in Section \ref{sect:model}, while the magnitude loss $m_{a}(i)$ is from Eq.~\ref{eq6} by giving sky brightness in left figure. From left to right, three equivalent error distributions are respectively calculated from sky brightness of peak, $3\sigma$ cut, and maximum. }}
  \label{fig3}
  \end{minipage}%
\end{figure}

Both items $m_{sky}^{'}$ and $m_{a}(i)$ in the Eq.~\ref{eq6} are unknown, while $I(i)$  is calculated by right side of Eq.~\ref{eq6}. The histogram of $I(i)$  is plotted in Fig.~\ref{fig2}. Considering existing of many random position errors and the large number of fibers targeting to point sources in selected exposures ($>3000$ for some exposures), it is reasonable to assume that at least a part of fibers have $m_{a}$ close to zero, that is, the faintest part on this histogram. We are able to estimate the sky brightness,  $m_{sky}^{'}$ on the the histogram \ref{fig2}. Then it is easy to determine the corresponding $m_{a}(i)=m_{sky}^{'}-I(i)$ for each individual fiber, and to solve the value of equivalent position error from the model described in Section 3.  

Fig.\ref{fig3} shows the fiber equivalent position error distributions of one exposure, calculated from three specified sky brightness values: the peak, the maximum (the outlet points manually rejected, which is discussed in the section \ref{subsect:error_sources}), and the value at $3\sigma$ cut, where $\sigma$ is the variation of the half Gaussian of right side from the peak of the impled sky brightness distribution in Fig.\ref{fig2}. The equivalent position error distribution from the peak sky brightness implies the working condition of fiber position units, and ignores the systematic errors that mainly contributed by the integration of guiding motion and the defocusing of whole focal plane. It represents the position errors caused by orthogonal mismatching, the random defocusing shifting from the focal plane, and the tilt (telecentric alignment error) of individual fibers. The result from $3\sigma$ cut is close to that from the maximum, and is convenient to exclude the outlets in the impled sky brightness distribution. The shifting between the distribution from the peak sky brightness and other two, $0.5'' \sim 0.7''$ in Fig.\ref{fig3}, can be considered as the value of systematic position error from the guiding motion and the whole focal surface defocusing.  

\begin{figure}[h]
  \begin{minipage}[t]{0.495\linewidth}
  \centering
   \includegraphics[width=90mm,height=75mm]{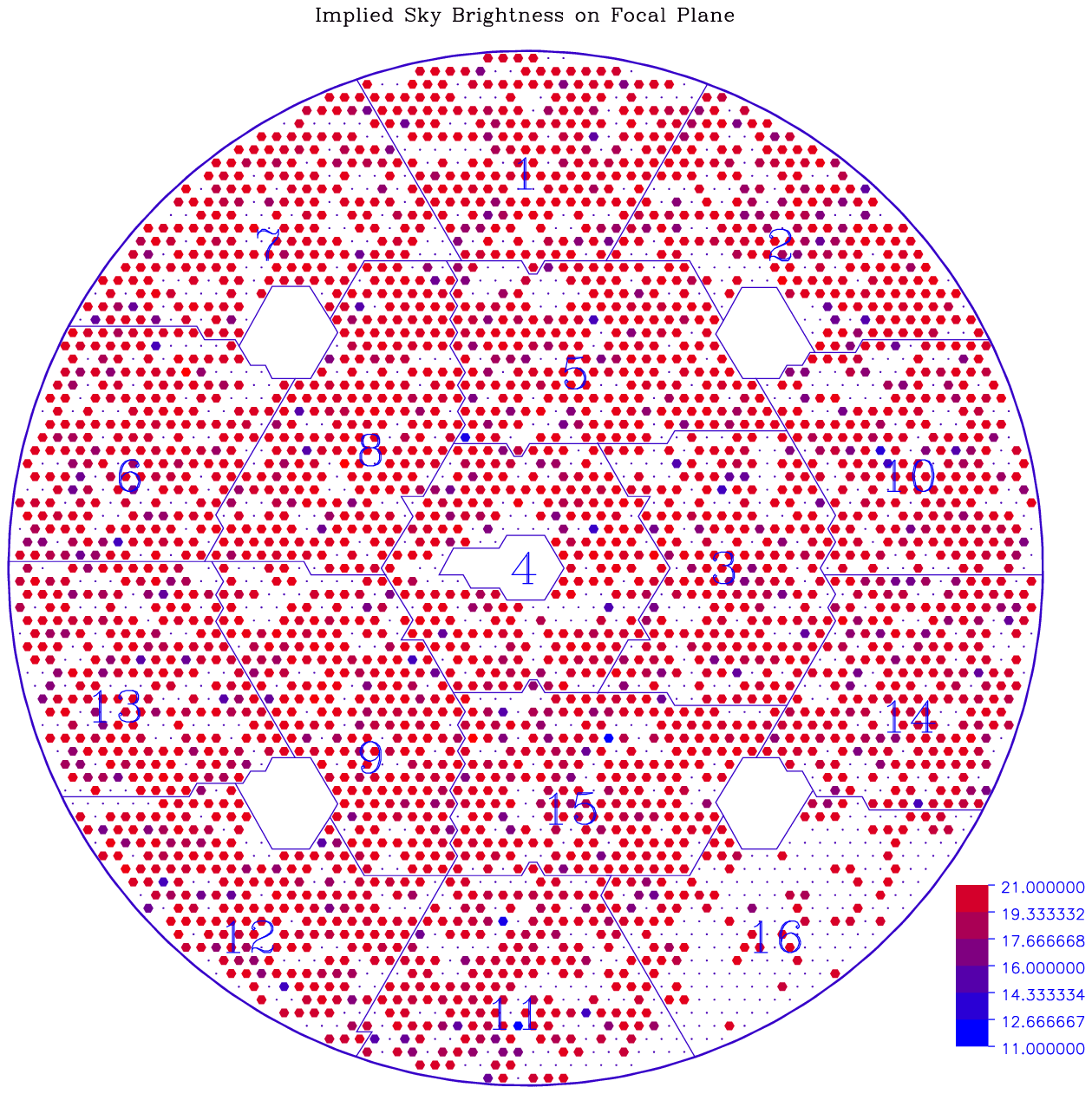}
  \end{minipage}%
  \begin{minipage}[t]{0.495\textwidth}
  \centering
   \includegraphics[width=90mm,height=75mm]{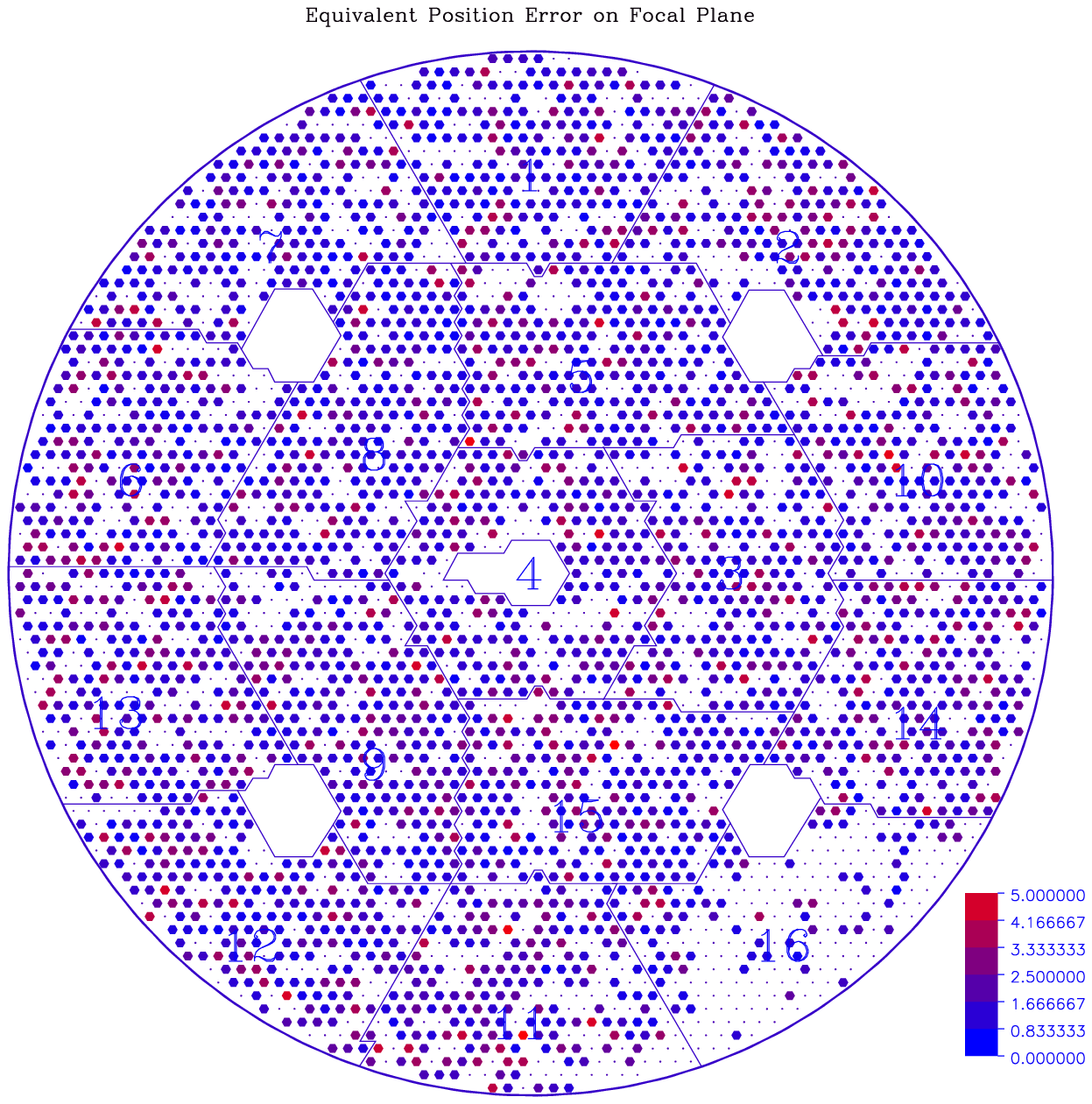}
  \end{minipage}%
  \caption{{\small The implied sky brightness distribution on focal plane (left), and the equivalent position error distribution on focal plane (right). The blank fiber unites were not assigned to point targets.}}
  \label{fig4}
\end{figure}

Fig.\ref{fig4} gives plots of the distributions of impled sky brightness and equivalent position error on the focal plane. At least for this exposure, there is no conspicuous evidence for a gradient across the focal plane.

\section{Discussion}
\label{sect:discussion}

\subsection{The Width of Point Spread Function (PSF), $W$}
\label{subsect:seeing}
As mantioned before, the width of PSF, $W$, is given by measuring FWHM of star spots on the guiding CCD images. A constant $W$ is assumed across the focal plane. This assumption can be checked by measuring the variation of W among images from four guiding cameras, which are mounted in a square on the focal plane.

In this technique, $W$ variation is counted as misfocusing, no matter individually or sysmatically. So the effects of $W$ variation contributes to equivalent position error if it is casued by individule fiber misfoucsing or the tilt and misfocusing of the focal plane. But the result could be misleading if it is caused by the local thermal disturbance on the forcal plane. 

Increasing the value of $W$, will flatten the equivalent position error distribution and expand the range of the distribution to the larger error side.

Another notable aspect about $W$ is the guiding motion. Following the point of view of Newman (~\cite{newman02}), we considered the effect on the flux loss due to the integration of guiding motion as a part of  position errors in this paper, and as main contributor to the systematic position errors. From another point of view, this effect also can be considered as an extended seeing disk during the exposure. The stacked guiding images could be used to measure this extended $W$.

\subsection{Influence of Atmospheric Transmittance}
\label{subsect:atmos}
Since not all components of sky light are from outside of earth atmosphere, the integrated atmospheric thickness passed by sky light is always less than that passed by targets light. So $\Delta m_{atm}$ depends on atmospheric thickness, too. The value of $\Delta m_{atm}$ is hard to measure in practice, because either the intensity of night sky components or the atmospheric transmittance continually varies during the exposure. While we can treat items of $m_{sky}(i)$ and $\Delta m_{atm}(i)$ as a whole, $m_{sky}^{'}(i)$ in Eq.\ref{eq4}, only the uniformity of both is required for Eq.\ref{eq6}.  

The major inside components of sky light include airglow, aurora, and pollution lights~(Gustafson, B., et al.~\cite{Gustafson07}). The aurora is weak at Xinglong, of which the latitude is about $40\dg N$. Both aurora and airglow are from the top level of atmosphere at an altitude of about 100 km and higher, therefore they have similar effective atmospheric thickness to that of targets light. So the major contributor to $\Delta m_{atm}$ is the artificial pollution lights from ground. Ice crystals and water droplets in clouds attenuate the light from outside atmosphere and reflect pollution from ground (Burke, D., et al. ~\cite{Burke10}), and ruin the uniformity of $m_{sky}$ and  $\Delta m_{atm}$. 

\subsection{Error Sources of Measurement}
\label{subsect:error_sources}

Uniformity of both sky brightness and atmospheric transmittance is essentially required for this method to measure fiber position error. So the telescope pointing of selected exposure is limited to angle close to zenith, e.g. zenith angle less than 30\dg. The clear, cloudless exposure condition is needed. Moonless is not necessary but the telescope pointing needs to keep distance to the moon in order to avoid the brightness gradient caused by the moon. For an exposure at larger zenith angle $>50\dg$, the error caused by gradient in $m_{a}$ could be larger than $0.1 mag$. It mainly affects the fibers with the equivalent position error less than 1.5$''$ for small seeing condition ($W$) in Fig.\ref{Fig1}. For the fibers having larger equivalent position error, the final results are not sensitive to this gradient, at least for the exposures with small seeing disks.

The stray lights, undetected cosmic ray, and unmasked warm column on CCD images contaminate the target spectrum, and lead to the solved sky brightness unusually faint. Actually, a few fibers meet this kind of situation in the bottom right corner of Fig.\ref{fig2}, which have isolated sky brightness $m_{sky}^{'} - m_{a}(i)$ value of about 22. These fibers are rejected while deciding the right edge of the histogram.

The assumption that at least a part of fibers with faintest sky brightness satisfies $m_{a}(i)=0$ may not be true. If so, the histogram in Fig.\ref{fig3} would shift leftwards. But the profile of the equivalent position error distribution changes little, and the order of fiber sorted by the value of position error remains unchanged. 

The precision of data reduction affects the measuring accuracy, especially, the precision of sky subtraction. 

\subsection{Compare with SNR and CCD Photographic Measurement}
\label{subsect:snr}
Position errors greatly affect SNR of observed spectra. Practically, the spectral SNR is often used as an indicator to fiber positioning errors. Newman (~\cite{newman02}) acknowledges that the real-time evaluation of the spectral SNR is a tool to compensate the variation in position errors. SNR is also used for statistically picking out the fiber units with large errors. Besides position errors, SNR affected by many other factors, such as vignetting, efficiency variation among spectrographs and CCD cameras, throughput variation among fibers, etc. Equivalent position error measurement overcome these aspects, because as the reference, the sky light goes through exact same aperture as targets light, and identically convolves the telescope and instrument response. From Fig.\ref{fig3}, this method implies the seperation of systematic and random errors.

Equivalent position error measurement and photographic measurement are complementary. The former measures the total position error, but is unable to distinguish the error sources. Photographic measurement could help to estimate the source by providing the orthogonal error component, which benefits the trouble shooting and solving for the fiber units with large position errors.  

This equivalent position error method is an instrument independent measurement. It could easily be applied to spectral data of other multi-fiber telescope to measure the total position errors for individual fibers.

\normalem
\begin{acknowledgements}
The authors would like to thank Prof. Hu Jingyao Prof. Wang Gang for the helpful discussions. We are grateful to the anonymous referee for valuable comments and suggestions.
Guoshoujing Telescope (the Large Sky Area Multi-Object Fiber Spectroscopic Telescope LAMOST) is
a National Major Scientific Project built by the Chinese Academy of Sciences. Funding for the project has
been provided by the National Development and Reform Commission. LAMOST is operated and managed
by the National Astronomical Observatories, Chinese Academy of Sciences.

\end{acknowledgements}
  

\begin{thebibliography}{99}

  \bibitem[1988]{Brodie88} Brodie, J., Lampton,M., Bowyer, S., 1988, \aj, 96, 2005

  \bibitem[2010]{Burke10} Burke, D., Axelrod, T. et al. 2010, \apj, 720, 811


  \bibitem[1996]{Chromey96} Chromey, F., Hasselbacher, D., 1996, \pasp, 108, 944

  \bibitem[2012]{cui12} Cui, X., Zhao,Y., et al., 2012, \raa, 12, 1197

  \bibitem[1989]{Donnelly89} Donnelly, R., Brodie, J., et al., 1989, \pasp, 101, 1046

  \bibitem[2012]{gu12} Gu, Y., Jin, Y., Zhai, C., 2012,Proceedings of Society of Photo-Optical Instrumentation Engineers (SPIE), 8450, ed. R. Navarro, 84503E, 9 pp

  \bibitem[2007]{Gustafson07} Gustafson, B., et al., 2007, IAU Transactions,  Vol. 26A, ed. O. Engvold. Cambridge: Cambridge University Press, 2007., pp.161-166

  \bibitem[2002]{newman02} Newman, P., 2002, \pasp, 114, 918

  \bibitem[1964]{Roach64} Roach, F., 1964, Space Science Reviews, 3, 512

  \bibitem[2012]{wang12} Wang, M., Zhao, Y., and Luo, A., 2012, Proceedings of Society of Photo-Optical Instrumentation Engineers (SPIE), 8450, ed. R. Navarro, 84503D, 10 pp
  
  \bibitem[1998]{xing98} Xing, X., Zhai, C,. et al., 1998, Proceedings of Society of Photo-Optical Instrumentation Engineers (SPIE), 3352, ed. L. Stepp, 839

  \bibitem[2012]{zhao12} Zhao, G., Zhao, Y., et al., 2012, \raa, 12, 723

\end{thebibliography}

\end{document}